\documentclass[pre, aps, floatfix, showpacs, twocolumn]{revtex4}
\usepackage{latexsym}
\usepackage{amssymb}
\usepackage{amsmath}
\usepackage{graphicx}
\usepackage{epsfig}
\usepackage{subfigure}

\newcommand{\stl}[1]{\mbox{$ \hspace{0.1em}
      \stackrel{\rule{0.4pt}{0.275ex}\hspace{0.40em} \!\!\!
      \overline{\hspace{0.06em}\vphantom{\rule{0.4pt}{0.0ex}}
      \hphantom{\mbox{$\displaystyle #1$}}
      \hspace{0.06em}  } \!\!\!\hspace{0.40em}\rule{0.4pt}{0.275ex}}
      {#1}\hspace{0.2em}$}}

\begin{document}
\title{Fluctuating Nematodynamics using the Stochastic Method of Lines }
\author{A. K. Bhattacharjee}
\author{Gautam I. Menon}
\author{ R. Adhikari}
\affiliation{The Institute of Mathematical Sciences,
C.I.T. Campus, Taramani, Chennai 600013, India}
\date{\today}

\begin{abstract}
We construct Langevin equations describing the  fluctuations of the tensor order
parameter $Q_{\alpha\beta}$ in nematic liquid crystals by adding noise terms to 
time-dependent variational equations that follow from the Ginzburg-Landau-de Gennes 
free energy. The noise is required to preserve the symmetry and tracelessness of 
the tensor order parameter and must satisfy a fluctuation-dissipation relation 
at thermal equilibrium. We construct a noise with these properties in a basis of 
symmetric traceless matrices and show that the Langevin equations can be solved 
numerically in this basis using a stochastic version of the method of lines. 
The numerical method is validated by comparing equilibrium probability 
distributions, structure factors and dynamic correlations obtained from these 
numerical solutions with analytic predictions.  We demonstrate excellent 
agreement between numerics and theory. This methodology can be applied to the 
study of phenomena where fluctuations in both the magnitude  and direction of 
nematic order are important, as for instance in the nematic swarms 
which produce enhanced opalescence  near the isotropic-nematic transition or the problem of 
nucleation  of the nematic from the isotropic phase. 
\end{abstract}
\date{\today}
\maketitle

\section{\bf Introduction}
Fluctuation phenomena in nematic liquid crystals are typically studied within
Ericksen-Leslie theory, which assumes that  the orientation of the normalized nematic director is the only 
fluctuating variable\cite{degenpro}. This approximation is adequate deep within the nematic phase, 
where the strength of nematic order is not significantly affected by thermal 
fluctuations. However, in the vicinity of the weakly first-order isotropic-nematic transition, significant 
fluctuations in the nematic order are observed, suggesting that the phase-only approximation
embodied in Leslie-Ericksen theory is inadequate \cite{ajp2007}. The study of nucleation in quenches from the isotropic to the nematic 
phase involves the growth of one phase within another, mandating the use of descriptions
capable of describing both isotropic and nematic phases on the same footing.  In these
and similar situations,  a tensorial description of nematic order which uses the symmetric, traceless 
quadrupole moment  tensor $Q_{\alpha\beta}$,  is appropriate, as first clarified by de Gennes in his
Ginzburg-Landau theory of the isotropic-nematic transition\cite{degennes}. The Ginzburg-Landau-de Gennes (GLdG) approach
provides a simple, but accurate  phenomenological description of nematic fluctuations in the
static case\cite{gralondej}. 

The description of the fluctuating dynamics of the orientation tensor within GLdG theory has received 
considerably less  attention\cite{stratv, olmgol}. Understanding the results of inelastic scattering experiments on
nematic systems\cite{bernepecora},  the description of the rate of nucleation into the nematic phase\cite{cutoedijks}, the
modeling of nontrivial stresses arising
from Casimir  interactions\cite{ajdaripelitiprost} and the calculation of the spectrum of capillary waves on the 
isotropic-nematic interface\cite{schgersch} 
are all problems which require a dynamical theory of fluctuations in the orientation tensor.  This problem 
is addressed in this paper, in which 
we present and solve the Langevin equations for dynamical fluctuations at 
equilibrium for the nematic orientation tensor. These are stochastic non-linear 
partial differential equations for the five components of the orientation tensor. 
Analytical solutions can be obtained when these equations are linearized. For the solution of the
general non-linear equations, we 
propose an efficient numerical method, based on a stochastic generalization of 
the method of lines. We compare our results with analytic results where such
calculations are possible,  finding excellent agreement.

\section{\bf Fluctuating nematodynamics}
\label{sec:2}
Orientational order in the nematic phase is described by a second-rank, 
symmetric traceless tensor $Q_{\alpha\beta}({\bf x}, t)$. This is the second 
moment of the microscopic orientational distribution function. The tensor can 
be expanded as 
\begin{equation}
\label{Qtensor}
Q_{\alpha \beta}=\frac{3}{2}S(n_{\alpha}n_{\beta} - \frac{1}{3}\delta_{\alpha \beta}) + 
\frac{1}{2}T(l_{\alpha}l_{\beta} - m_{\alpha}m_{\beta}).
\end{equation}
The three principal axes of this tensor, obtained by diagonalizing $Q_{\alpha\beta}$ 
in a local frame, specify the direction of nematic ordering ${\bf n}$, the 
codirector ${\bf l}$ and the joint normal to these, labeled by ${\bf m}$. The principal 
values $S$ and $T$ represent the strength of ordering in the direction of 
${\bf n}$ and ${\bf m}$, quantifying, respectively, the degree of uniaxial 
and biaxial nematic order. 

The static fluctuations of $Q_{\alpha\beta}$ can be calculated 
from a Ginzburg-Landau functional, first proposed by de Gennes, based on an 
expansion in rotationally invariant combinations of $Q_{\alpha\beta}$ 
and its gradients. The Ginzburg-Landau-de Gennes functional 
$F$ is
\begin{eqnarray}
\label{frener}
F &=& \int d^3{\bf x} [\frac{1}{2}ATr{\bf Q}^{2} + \frac{1}{3}BTr{\bf Q}^{3} + 
\frac{1}{4}C(Tr{\bf Q}^{2})^{2} \\\nonumber
&+& E^{\prime}(Tr{\bf Q}^{3})^{2} + \frac{1}{2}L_{1}(\partial_{\alpha}Q_{\beta\gamma})(\partial_{\alpha}Q_{\beta\gamma})]. 
\end{eqnarray} 
Here, $A = A_{0}(1 - T/T^{*})$ $T^{*}$ denoting the supercooling transition 
temperature, $A_{0}$ a constant, $L_{1}$  is an elastic 
constant and $\alpha, \beta, \gamma$ denote the 
Cartesian directions. From the inequality $\frac{1}{6}(Tr{\bf Q}^{2})^{3} \geq (Tr {\bf Q}^{3})^{2}$, higher powers 
of $Tr{\bf Q}^{3}$ can be excluded for the description of the uniaxial phase. 
Uniaxial phases are described by $E^{\prime}$ = 0 while biaxial phases 
require $E^{\prime} \neq 0$. For the nematic phase (rod-like molecules) 
$B<0$ whereas for the discotic phase (plate-like molecules) $B>0$. The 
quantities C and $E^{\prime}$ must always be positive to ensure boundedness 
and stability of the free energy in all phases. We omit other symmetry-allowed gradient terms in this
paper, thus   working in the limit where all three Frank constants are assumed to be
equal. Such symmetry-allowed terms, as also total derivative surface terms,  can be accounted for
without essential  change, using the numerical method  described below.

In the limit that hydrodynamic interactions may be neglected, {\it i.e.} the Rouse
or free-draining limit,  the dynamical fluctuations of $Q_{\alpha\beta}$ are not coupled to 
other hydrodynamic variables. The Langevin equations are those appropriate to
a non-conserved order parameter with an overdamped, 
relaxational dynamics of the form 
\begin{equation}
\label{Qdynamics}
\partial_t Q_{\alpha\beta} = - \Gamma_{\alpha
\beta\mu\nu}{\delta F\over\delta Q_{\mu\nu}} + \xi_{\alpha\beta},
\end{equation}
Here the kinetic coefficients $\Gamma_{\alpha\beta\mu\nu}$, defined as 
$\Gamma_{\alpha\beta\mu\nu} = \Gamma[\delta_{\alpha\mu}\delta_{\beta\nu} + 
\delta_{\alpha\nu}\delta_{\beta\mu} - \frac{2}{3}\delta_{\alpha\beta}
\delta_{\mu\nu}]$, ensure that the dynamics preserves the symmetry and 
tracelessness property of the order parameter. In the absence of long-range 
forces, a local approximation for the kinetic coefficients is adequate
and $\Gamma$ can be taken as constant. The $\xi_{\alpha\beta}$ are symmetric, traceless
Gaussian white noises, which satisfy a fluctuation-dissipation relation 
at equilibrium of the form
\begin{eqnarray}
\langle \xi_{\alpha\beta}({\bf x}, t) \rangle &=& 0, \\
\langle \xi_{\alpha\beta}({\bf x}, t) \xi_{\mu\nu}({\bf x^{\prime}},
t^{\prime})\rangle &=& 2k_BT\Gamma_{\alpha\beta\mu\nu} \delta({\bf x - 
x^{\prime}})\delta(t - t^{\prime}). 
\end{eqnarray}
Here $k_B$ is the Boltzmann constant, $T$ the temperature and $\langle \rangle$ denotes 
the average over the probability distribution of the noise.  These Langevin equations, together with the fluctuation-dissipation relation for the 
noise, ensure that the  stationary one-point probability distribution  of $Q_{\alpha\beta}$, $P[Q_{\alpha\beta}]$,  converges to Boltzmann equilibrium with $P[Q_{\alpha\beta}] \sim \exp(-F/k_BT)$.

The equations above are five coupled, non-linear stochastic partial differential equations, with a noise term which has
a tensorial structure. A numerical method of solution must maintain the symmetry and traceless of $Q_{\alpha\beta}$.
To ensure equilibrium dynamics, it must  also maintain the balance between fluctuation and dissipation. These two stringent requirements 
may be satisfied by transforming to a basis in which $Q_{\alpha\beta}$ is traceless and symmetric by construction. Symmetry and tracelessness of $Q_{\alpha\beta}$ is automatic. As we show below, the  noise can be constructed out of independent Gaussian
noises. 

We expand the orientational tensor in a basis of symmetric traceless 
matrices $T^{i}_{\alpha\beta}$  as 
\begin{equation}
Q_{\alpha\beta}({\bf x}, t) = \sum_{i=1}^{5}a_{i}({\bf x}, t)T^{i}_{\alpha\beta}
\end{equation}
with ${\bf T}^{1} = \sqrt{3/2}\, \stl{{\bf \hat{z}\hat{z}}}, {\bf T}^{2} =  
\sqrt{1/2}\, ({\bf \hat{x} \; \hat{x} - \hat{y} \; \hat{y}}), {\bf T}^{3} = 
\sqrt{2}\; \stl{{\bf \hat{x} \; \hat{y}}}, {\bf T}^{4} = \sqrt{2}\; \stl{{\bf 
\hat{x} \; \hat{z}}}$ and ${\bf T}^{5} = \sqrt{2}\; \stl{{\bf \hat{y} \; \hat{z}}}$. 
The complete basis of matrices is orthogonal in the sense that $T^{i}_
{\alpha\beta}T^{j}_{\alpha\beta} = \delta_{ij}$. In previous work we have presented 
explicitly the equations for the basis coefficients $a_i({\bf x}, t)$ that follow 
from the deterministic part of the relaxational kinetics $\partial_t Q_{\alpha\beta} 
= - \Gamma_{\alpha\beta\mu\nu} \:{\delta F/\delta Q_{\mu\nu}}$ \cite{bhmead}. (These
differ from the equations derived by others in that  we include all non-linearities 
as well as an additional symmetry-allowed gradient ($L_2)$ term.) Here
we focus on how an explicit construction of the 
noise can be implemented by expanding in the same basis. 

We expand the noise as
\begin{equation}
\xi_{\alpha\beta}({\bf x}, t) = \sum_{i=1}^{5}\xi_{i}({\bf x}, t)T^{i}_{\alpha\beta},
\end{equation}
where each $\xi_i({\bf x},t)$ is a zero-mean Gaussian white noise. From the 
orthogonality of the basis the inverse relation is
\begin{equation}
\xi_{i}({\bf x}, t) = 
\sum_{\alpha,\beta}\xi_{\alpha\beta}({\bf x}, t)T^{i}_{\alpha\beta}.
\end{equation}
From this, and the fluctuation-dissipation relation it follows that
\begin{eqnarray}
\langle \xi_{i}({\bf x}, t)\xi_{j}({\bf x}^{\prime}, t^{\prime}) \rangle &=&
\sum_{\alpha\beta\mu\nu} \langle \xi_{\alpha\beta}({\bf x}, t)\xi_{\mu\nu}
({\bf x}^{\prime}, t^{\prime})\rangle T^{i}_{\alpha\beta}T^{j}_{\mu\nu}, \\ \nonumber
 &=&
\sum_{\alpha\beta\mu\nu} 2k_BT\Gamma_{\alpha\beta\mu\nu}  T^{i}_{\alpha\beta}
T^{j}_{\mu\nu}\delta({\bf x - x^{\prime}})\delta(t - t^{\prime}), \\ \nonumber
&=&
2k_BT\Gamma \delta_{ij}\delta({\bf x - x^{\prime}})\delta(t - t^{\prime}).
\end{eqnarray}
This shows that the non-trivially correlated noise $\xi_{\alpha\beta}$ can be 
constructed from uncorrelated noises $\xi_i$. Thus, by construction, the 
noise $\xi_{\alpha\beta}$ is symmetric, traceless and satisfies the 
fluctuation-dissipation relation.

When anharmonic terms are ignored in the Ginzburg-Landau-de Gennes functional, the Langevin equations are
linear and correlation functions can be calculated explicitly in the ${\bf T}$ basis. Then, 
the Langevin equations of motion in terms of 
\begin{equation}
a_i({\bf q}, t) = 
\int d^3x \exp(-i{\bf q\cdot x})a_i({\bf x}, t),
\end{equation}
 are
\begin{equation}
\label{nemlang}
\partial_t a_i({\bf q}, t) = -\Gamma(A + L_1 q^2)a_i({\bf q}, t) + \xi_i({\bf q}, t).
\end{equation}
From this, the static and dynamic correlations follow immediately,
\begin{equation}
C_{ij}({\bf q}) = \langle a_i({\bf q})a_j(-{\bf q}) \rangle = \frac{k_BT}{A + 
L_1 q^2}\delta_{ij},
\end{equation}
and
\begin{eqnarray}
C_{ij}({\bf q}, \tau) & =& \langle a_i({\bf q}, t)a_j(-{\bf q}, t + \tau) \rangle \\\nonumber
&=& C_{ij}({\bf q}) \exp[-\Gamma(A + L_1 q^2)\tau],
\end{eqnarray}
where $C_{ij}({\bf q})$ is the static structure factor.
The static and dynamic correlations for $Q_{\alpha\beta}$ are then
obtained by returning to the original basis. The stationary probability distribution 
generated by the Langevin dynamics is Gaussian with zero mean and variance 
$C_{ij}({\bf q})$, consistent with Boltzmann equilibrium.

\section{\bf Stochastic method of lines}
\label{sec:3}
The fluctuating nematodynamics equations contained in Eq.~(\ref{Qdynamics})  are five non-linear stochastic partial differential equations.
In general these have no analytical solutions and reliable numerical methods are therefore essential for their study. 
Here we combine the method of lines for solving initial-value partial differential equations with a stochastic Runge-Kutta integrator for systems of stochastic ordinary differential equations. This enables us  to construct an accurate and efficient solver for the equations of fluctuating nematodynamics. Our
results here  build on previous work \cite{bhmead}, where a method of lines approach was used to  solve the deterministic time-dependent Ginzburg-Landau 
equations numerically. The methodology here can thus be thought of as a generalization of the method of lines to stochastic partial differential equations. 

\begin{figure}
\centering
\includegraphics[width=8cm,]{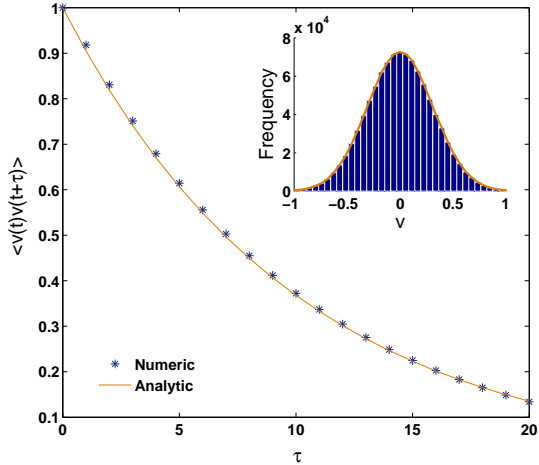}
\caption{(Color online) Autocorrelation function for the Ornstein-Uhlenbeck process, showing
$\langle v(t) v(t+\tau) \rangle$ as a function of the time increment $\tau$.
The inset shows the histogram of fluctuations. It is Gaussian with the expected variance. The numerical parameters  chosen are 
$\Gamma = k_BT = 0.1, dt = 1.0$. The average is taken over $10$ independent realizations
while the integration is performed for $10^6$ SRK4 steps.} 
\label{fig:OUvel}
\end{figure}

The method of lines is based on the idea of semidiscretisation, where an initial-value partial differential equation in space and time is  discretised only in the spatial variable \cite{liskovets}. This yields a (possibly large) system of ordinary differential equations  which is then solved by standard numerical integrators. To apply this method to stochastic partial differential equations, we must account for the fact that integrators for ordinary differential equations do not automatically provide efficient and accurate solutions of stochastic differential equations. Qualitatively, the noise term in a stochastic differential equation is a rapidly varying function and hence must  be integrated with some care. 
At a more technical level, the noise is a Wiener process and the theory of stochastic integration must be used to evaluate it correctly \cite{gardiner}. 

Common stochastic integrators include those due to Maryuama \cite{maruyama} and Milstein \cite{milstein}. In this work, we use an integrator proposed recently by Wilkie \cite{wilkie}, based on a multi-step Runge-Kutta strategy. The integrator is accurate  and easy to implement by making small changes to a deterministic Runge-Kutta integrator. Further, since it is an explicit integrator, no matrix inversions are involved. This makes it attractive when the method of lines discretisation produces a large system of ordinary differential equations, as  in our case. 

To test Wilkie's algorithm for a stochastic Runge-Kutta integrator, henceforth denoted as SRK4, we first check that the fluctuation-dissipation is obeyed.
We performed a simple benchmark test on the Ornstein-Uhlenbeck process, represented as the Ito differential equation,
\begin{equation}\label{vbrown}
dv(t) = -\Gamma v(t) dt + dW(t),
\end{equation} 
where $dW(t) = \sqrt{2k_BT\Gamma dt}\: \mathcal{N}(0, 1)$ is the increment of the 
stochastic variable in the interval $dt$, and $\mathcal{N}(0, 1)$ is a zero-mean unit-variance normal deviate. By construction, the increments of this stochastic variable are independent and 
normally distributed with mean $\langle dW(t) \rangle = 0$. The particular choice of the variance ensures that the equilibrium distribution of $v$ is a Gaussian with variance $k_BT$.  The stationary two-point autocorrelation of the velocity from Eqn.(\ref{vbrown}) is,
\begin{equation}
\langle v(t) v(t+\tau) \rangle = k_BT \texttt{exp}(-\Gamma\tau).
\end{equation}
Fig.(\ref{fig:OUvel}) shows the  autocorrelation as a function of time and the histogram of equal-time fluctuations of $v$. The variance $\langle v^2 \rangle = k_BT$ is correctly reproduced, as is the exponential decay of the autocorrelation function. We conclude that SRK4 is suitable as an integrator for problems where the fluctuation-dissipation relation must be maintained.

\section{\bf Numerical Method}
\label{sec:4}

\begin{figure}
\centering
\includegraphics[width=8cm]{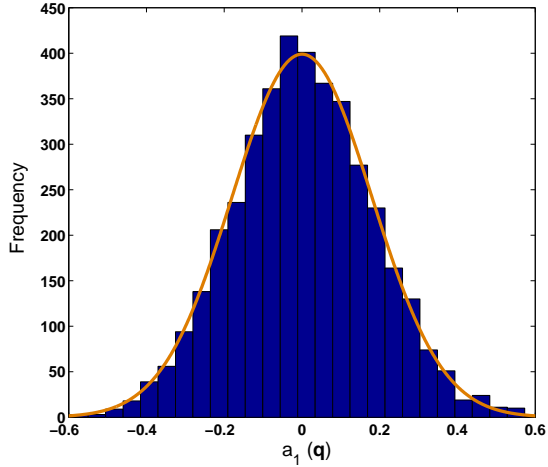}
\caption{(Color online) Histogram of the real part of $a_i({\bf q})$ for $n_x = n_y = 6$ in a box of dimension $L_x = L_y = 16$, using a harmonic free energy, with $k_BT = A = 0.05$ and $L_1 = 0.5$, $\Gamma = 1.0$. The fluctuations are Gaussian with the expected variance. The histogram is obtained from 20 independent realizations, each realization contributing 4000 time steps.}
\label{fig:probhist}
\end{figure}

\begin{figure}
\subfigure[]{\label{fig:Sq}
    \centering
    \includegraphics[width=8cm]{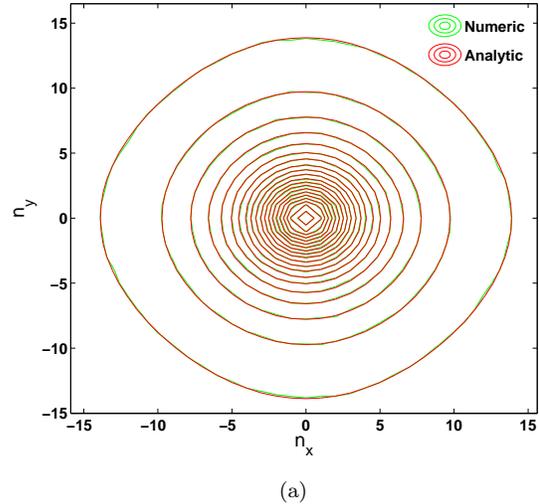}
}
\subfigure[]{\label{fig:angsq}
    \centering
    \includegraphics[width=8cm]{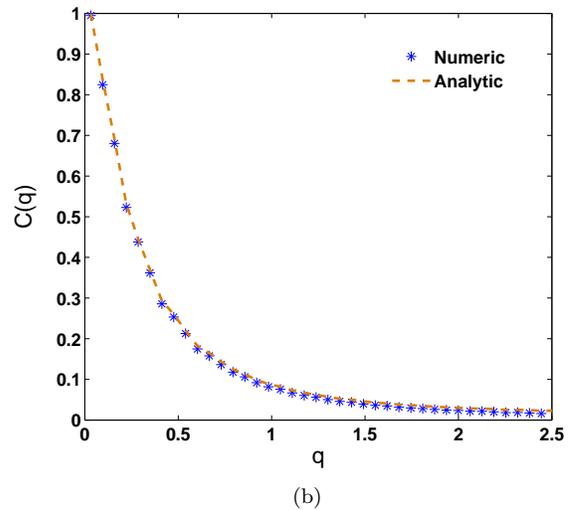}
}
\caption{(Color online) (a) Contour plot of the 
structure factor $C_{ij}({\bf q})$  and  (b) angular average of $C_{ij}({\bf q})$. The parameters 
are $k_BT = A = 0.05, L_1 = 0.5, \Gamma = 1.0$. The time averaging is over $5\times10^4$ time steps and ensemble averaging is over 20 independent realizations in a box with $L_x = L_y = 64$. The relaxation time scale is $\tau = (\Gamma A)^{-1} = 20$, the diffusion time scale of the smallest Fourier mode is $\tau_d = L_x^2/(4\pi^2 L_1 \Gamma) = 207.51$ and the correlation length is $\lambda = \sqrt{L_1/A} = 3.16$.} 
\end{figure}

We now apply the method of lines together with SRK4 to obtain a stochastic method lines discretisation (SMOL) for the equations of fluctuating
nematodynamics. We benchmark  our numerical results by comparing  autocorrelations within a harmonic theory which
accurately describes fluctuations about the isotropic phase. We then consider expansions about the ordered state, comparing static correlations obtained analytically within the Frank approximation with our numerical results.

\begin{figure}
\centering
\includegraphics[width=8cm]{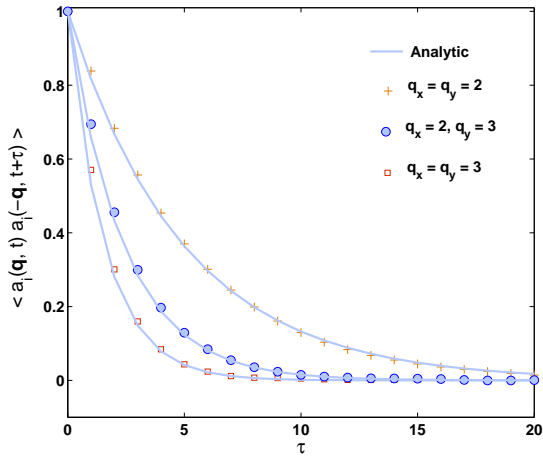}
\caption{(Color online) Autocorrelation $C_{ij}({\bf q},\tau)$ for the linear Langevin equation, calculated for small wavenumbers. Numerical parameters are $k_BT = A = 0.05,  L_1 = 
0.5, \Gamma = 1.0$. The integration is performed for $4.2\times10^3$ 
time steps on a $16^2$ grid. The time average is taken over $4\times10^3$ time steps 
and an ensemble average is taken over 40 independent realizations. The relaxation time scale 
is $\tau = (\Gamma A)^{-1}$ = 20, the diffusion time scale of the shortest Fourier mode 
$\tau_d = L_x^2/(4\pi^2 L_1 \Gamma) \sim 13$ and the correlation length $\lambda = \sqrt{L_1/A} = 
3.16$.}
\label{fig:autocor}
\end{figure}

\begin{figure}
\centering
\includegraphics[width=8cm]{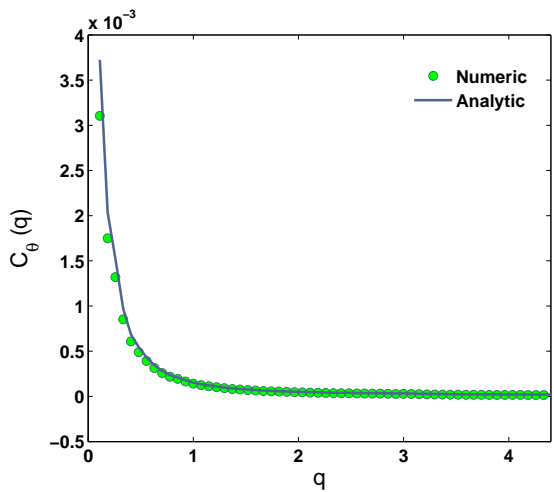}
\caption{(Color online) Angularly averaged structure factor of the nematic phase. The numerical  parameters are $k_BT = 0.05, A = -3.5, B = -10k_BT, C = 2.67, L_1 = 32.0, \Gamma = 0.01$ 
on a $64^2$ grid. The relaxation time scale is $\tau = (\Gamma A)^{-1}$ = 28.57, the diffusion time scale of the shortest Fourier mode is $\tau_d = L_x^2/(4\pi^2 L_1 \Gamma) = 324.23$ and the correlation length is $\lambda = \sqrt{L_1/A} = 3.02$. The time average is taken over $5 \times 10^{4}$ time steps and an ensemble average is 
taken over $20$ independent realizations.}
\label{fig:thetqsqav}
\end{figure}

We use a finite-difference discretisation with nearest-neighbour stencils for gradients and the Laplacian.  We implement periodic boundary conditions.
Specifically, in three dimensions, we consider a box of dimension  $L_x, L_y$ and $L_z$  along the Cartesian directions, and grid these
lengths with equal grid spacing $\Delta x = \Delta y = \Delta z = 1$. The latter defines lattice units for the spatial coordinate. We define corresponding discrete
time units for the temporal variables by choosing $\Delta t = 1$. Fourier modes are labelled by the wave-vector ${\bf q} = (q_x, q_y, q_z)$, where each component is of the form $q_{\alpha} = {2\pi n_{\alpha}}/{L_{\alpha}}$, with $n_{\alpha} = 0,1,2,\ldots,(L_{\alpha}-1)$. 

With this discretisation, the Laplacian in Fourier space is given by
\begin{equation}
\mathcal{L}({\bf q})  = 2[\cos(q_x) + \cos(q_y) + \cos(q_z) -3].
\end{equation}
The nearest-neighbour finite difference stencil suffers from lack of isotropy at high wavenumbers. This can be improved through the use of
higher-point stencils \cite{abramstegun, pakar, shioon}. 

Applying the method of lines discretisation to Eq.~(\ref{Qdynamics})  reduces it to a system of stochastic ordinary differential equations, whose Fourier representation in the harmonic approximation of Eq.~(\ref{nemlang}) is
\begin{equation}
\label{discrnemlang}
\partial_t a_i({\bf q}, t) = -\Gamma\mathcal{D}a_i({\bf q}, t) + \xi_i({\bf q}, t).
\end{equation}
The Fourier representation of the drift-diffusion dynamics is encoded in the linear operator $\mathcal D$, 
\begin{equation}
\label{q2discr}
\mathcal{D} = A - L_1\mathcal{L}({\bf q}).
\end{equation}
Fourier representations of the one and two-dimensional method of lines discretisations are obtained by setting the corresponding wavenumbers to zero. The static and dynamic autocorrelations in Fourier space follow in a straightforward manner though the replacement of   $q^2$ by its discrete Laplacian representation. The results are
\begin{eqnarray}
C_{ij}({\bf q}) &=& \frac{k_BT}{\mathcal{D}}\delta_{ij}, \\
C_{ij}({\bf q},\tau) &=& C_{ij}({\bf q}) \exp(-\Gamma \mathcal{D}\tau). \\\nonumber
\end{eqnarray}
It is also useful to define an angle-averaged structure factor $C(q) = \sum_{|{\bf q}|=q} C_{ij}({\bf q}\rangle$ for comparison with the numerical simulation.

We now compare theoretical and numerical results: In Fig.~(\ref{fig:probhist}) we show the histogram of the $a_i({\bf q})$ for a particular Fourier mode. This is normally distributed, as expected, with zero mean and variance as required by thermal equilibrium. Similarly, all Fourier modes examined have correct normal distributions. The variances obtained are compared in Fig.~\ref{fig:Sq} with the analytical values by plotting contours of $C_{ij}({\bf q})$. There is excellent agreement. A close inspection reveals some degree of anisotropy in both the analytical and numerical results at high wavenumbers. This is attributed to the lack of isotropy of the nearest-neighbour finite-difference Laplacian mentioned earlier. However, the anisotropies  are removed upon angular averaging, as shown in 
 Fig.~\ref{fig:angsq}. Thus, the present discretisation  should be adequate in most cases, unless highly accurate isotropies are required 
 from the simulation. From these  results, we conclude that correlations in thermal equilibrium are accurately captured by the stochastic method of lines approach. 
 
We next compare the dynamics of fluctuations at equilibrium, by comparing  two-point autocorrelation functions calculated analytically and numerically. Fig.(\ref{fig:autocor}) shows $C_{ij}({\bf q},\tau)$ for three sets of Fourier modes. The exponential decay of the autocorrelation function is reproduced accurately within the numerics
and fit the theoretical curve very closely. We conclude, therefore, that the stochastic method of lines accurately reproduces both static and dynamic fluctuations in a harmonic theory. 

Finally, we compare theory and simulation in a situation where a linearization of the $Q_{\alpha \beta}$ equations about
$Q_{\alpha \beta} = 0$ is inapplicable, that of director fluctuations within the nematic phase.  
In the ${\bf T}$ basis, the equations of motion are
\begin{eqnarray}
\partial_{t}a_{i} &=& - \Gamma \;[(A + C TrQ^{2})a_{i} \\\nonumber
&+& (B +  6E^{\prime}TrQ^{3}) 
T_{\alpha\beta}^{i}\stl{{Q_{\alpha\beta}^{2}}} - L_{1}\nabla^{2}a_{i}] + \xi_i.
\end{eqnarray}
where, $\stl{{Q_{\alpha\beta}^{2}}}$ is the traceless symmetric projection of $Q^2_{\alpha\beta}$.These equations of motion are anharmonic, and the analytical solutions  obtained earlier within the
harmonic expansion are  no longer available for comparison. We therefore extract the fluctuations of the angular 
displacements from the uniform nematic ground state using an approach based on the Frank free energy.

Consider an uniform uniaxial nematic with director ${\bf n}_0$ with small fluctuations
$\delta {\bf n}({\bf x})$. Decomposing the fluctuations into parts parallel and perpendicular to ${\bf n}_0$ and  imposing the normalization of the director, 
we find from the Frank free energy that, 
\begin{equation}
\langle |\delta {\bf n}_\perp({\bf q})|^2\rangle = \frac{k_BT}{Kq^2}, 
\end{equation}
where $K = (9S^2/2)L_1$ is a Frank constant \cite{gralondej}. Since  fluctuations in the plane perpendicular to ${\bf n}_0$ can be characterized through a 
single angle $\theta$, an equivalent result is $\langle \theta({\bf q}) \theta(-{\bf q}) \rangle =  k_BT/Kq^2$. In the semidiscrete representation, 
we obtain
\begin{equation}
\langle \theta({\bf q}) \theta(-{\bf q}) \rangle = -\frac{k_BT}
{K\mathcal{L}({\bf q})}.
\end{equation}   

Fig.(\ref{fig:thetqsqav}) shows the angular average of the static correlations of director fluctuations
$C_{\theta}(q) = \sum_{|{\bf q}|=q} \langle\theta({\bf q}) \theta(-{\bf q})\rangle$. The formally divergent
${\bf q} = 0$ mode is excluded both from the numerical data and 
analytical result. Given that  the analytical result  is obtained from a linearization about the aligned state whereas the numerical solution
is calculated from the equations of motion arising from the full non-linear free energy, the agreement between theory and  simulation is 
satisfactory. The stochastic method of lines  thus  accurately captures equilibrium fluctuations in the ordered state as well as in the disordered one. 

\section{Discussion and conclusion}
The numerical method of solution presented in this paper can be applied to a variety of problems in nematodynamics where accounting for
fluctuations in nematic order are important. To the best of our knowledge, no systematic study of anharmonic fluctuations exists within the time-dependent Ginzburg-Landau framework. In previous work, Stratonovich \cite{stratv} presented fluctuating equations of motion for harmonic fluctuations in terms of Langevin equations, analyzing
these equations within the traceless symmetric basis  described here in a straightforward manner\cite{olmgol}. Our equations of motion contain
the necessary nonlinearities and our numerical methodology accounts for them  in a computationally straightforward way. 

We also present, to the best of our knowledge for the first time,  a systematic stochastic integration scheme capable of yielding highly accurate solutions of the  non-linear equations of nematodynamics. These
equations  can be applied to study fluctuations of nematic order at the isotropic-nematic interface, pseudo-Casimir interactions close to the isotropic-nematic transition, as well as nucleation phenomena in nematogens. The study of  these and similar  problems is ongoing and will be reported elsewhere.  

\section{\bf Acknowledgements}
GIM thanks the DST(India) and the Indo-French Centre for the Promotion of Advanced Research (CEFIPRA) for support. 

\bibliography{references}
\end{document}